\documentclass{aa}
\usepackage[varg]{txfonts}
\usepackage{graphicx}
\usepackage{supertabular,booktabs}
\usepackage{subfig}
\usepackage{natbib}
\bibpunct{(}{)}{;}{a}{}{,}

\begin{document}

 \title{Spectroscopic mapping of the physical properties of supernova remnant N\,49\thanks{Based on observations obtained at the Southern Astrophysical Research (SOAR) telescope, which is a joint project of the Minist\'{e}rio da Ci\^{e}ncia, Tecnologia, e Inova\c{c}\~{a}o (MCTI) da Rep\'{u}blica Federativa do Brasil, the U.S. National Optical Astronomy Observatory (NOAO), the University of North Carolina at Chapel Hill (UNC), and Michigan State University (MSU).} }

\author{D. Pauletti
\and M. V. F. Copetti} 

\institute{Laborat\'orio de An\'alise Num\'erica e Astrof\'isica, Departamento de Matem\'atica; Programa de P\'os-Gradua\c{c}\~ao em F\'isica, Universidade Federal de Santa Maria, 97119-900, Santa Maria, RS, Brazil}

\date{Received date / Accepted date }

\abstract
{Physical conditions inside a supernova remnant can vary significantly between different positions. However, typical observational data of supernova remnants are integrated data or contemplate specific portions of the remnant.}
{We study the spatial variation in the physical properties of the N\,49 supernova remnant based on a spectroscopic mapping of the whole nebula.}
{Long-slit spectra were obtained with the slit ($\sim4\arcmin \times 1.03\arcsec$) aligned along the east-west direction from 29 different positions spaced by $2\arcsec$ in declination. A total of 3248 1D spectra were extracted from sections of  $2\arcsec$ of the 2D spectra. More than 60 emission lines in the range 3550\,\AA{} to 8920\,\AA{} were measured in these spectra. Maps of the fluxes and of intensity ratios of these emission lines were built with a spatial resolution of $2\arcsec \times 2\arcsec$.}
{An electron density map has been obtained using the [S\,{\sc ii}]\,$\lambda6716/\lambda6731$ line ratio. Values vary from  $\sim$500\,cm$^{-3}$ at the northeast region to more than 3500 cm$^{-3}$ at the southeast border. We calculated electron temperature using line ratio sensors for the ions S$^{+}$, O$^{++}$, O$^{+}$, and N$^{+}$. Values are about 3.6$\times10^{4}$\,K for the O$^{++}$ sensor and about 1.1$\times10^{4}$\,K for other sensors. The H$\alpha$/H$\beta$ ratio map presents a ring structure with higher values that may result from collisional excitation of hydrogen. We detected an area with high values of  [N\,{\sc ii}]\,$\lambda6583$/H$\alpha$ extending from the remnant center to its northeastern border, that can be indicating an overabundance of nitrogen in the area due to contamination by the progenitor star. We found a radial dependence in many line intensity ratio maps. We observed an increase toward the remnant borders of the intensity ratio of any two lines in which the numerator comes before in the sequence [O\,{\sc iii}]\,$\lambda5007$, [O\,{\sc iii}]\,$\lambda4363$, [Ar\,{\sc iii}]\,$\lambda7136$, [Ne\,{\sc iii}]\,$\lambda3869$, [O\,{\sc ii}]\,$\lambda7325$, [O\,{\sc ii}]\,$\lambda3727$, He\,{\sc ii}\,$\lambda4686$, H$\beta$\,$\lambda4861$, [N\,{\sc ii}]\,$\lambda6583$, He\,{\sc i}\,$\lambda6678$, [S\,{\sc ii}]\,$\lambda6731$, [S\,{\sc ii}]\,$\lambda6716$, [O\,{\sc i}]\,$\lambda6300$, [Ca\,{\sc ii}]\,$\lambda7291$, Ca\,{\sc ii}\,$\lambda3934$, and [N\,{\sc i}]\,$\lambda5199$.}
{}


\maketitle


\section{Introduction}

Supernova remnants (SNRs) commonly have inhomogeneous structures. Their morphology does not always follow the spherical nebula radial symmetries predicted by basic models for SNR evolution \citep{1972ARA&A..10..129W,1977ARA&A..15..175C}. This is a result of the high influence of interstellar medium (ISM) characteristics (such as density and temperature fluctuations, and magnetic field intensity and direction) on the SNR structure. Nevertheless, these characteristics are not the aim of most observational studies on SNRs, which typically present integrated spectra or spectra from only some specific positions in the remnant.

The brightest optical SNR in the Large Magellanic Cloud (LMC), N\,49 (LHA 120-N\,49; SNR B0525-66.1), is an ideal object for many types of scientific approaches on SNRs. However, only a few of them include spatially resolved optical observations that thoroughly cover the object. A first line emission mapping of N\,49 was obtained by \citet{1979ApJ...231L.147D} for the H$\beta$, [Fe\,{\sc xiv}]\,$\lambda$5303, [O\,{\sc iii}]\,$\lambda$5007, and  [N\,{\sc ii}]\,$\lambda$6583 lines. Their analysis of the relative intensities between [Fe\,{\sc xiv}] and H$\beta$ revealed a density gradient in N\,49 and has reinforced the cloudlet structure interpretation for SNRs proposed by \citet{1978ApJ...219L..23M}.

\citet{1992ApJ...394..158V} obtained optical CCD images of N\,49 using interference filters in [O\,{\sc iii}]\,$\lambda$5007, [O\,{\sc i}]\,$\lambda$6300, [S\,{\sc ii}]\,$\lambda\lambda$6716, 6731, and H$\alpha$ lines. A spatial displacement between low- and high-ionization line emission was revealed by an increase in the [O\,{\sc iii}]\,$\lambda$5007 line ratio to both [O\,{\sc i}]\,$\lambda$6300 or  H$\alpha$ in the object borders. Remarkably, the remnant optical shell that apparently was incomplete when seen in the H$\alpha$ light became well defined in this ratio maps, in agreement with the remnant morphology as observed at infrared and X-rays bands \citep{1995ApJ...448..623D,1999ApJS..123..467W}. The authors also derived electron density and temperature from long-slit spectroscopy data, but only for sparse locations.

The high [O\,{\sc iii}]\,$\lambda$5007/H$\alpha$ values in the remnant borders were also observed by \citet{2007AJ....134.2308B} in their photometric maps. The authors suggested that this might be caused by an offset in the [O\,{\sc iii}] and H$\alpha$ emission peaks, a consequence of a delay between the emissions of these two lines as shocks interact with a clumpy medium. This was previously suggested by \citet{1983ApJ...275..636R} to explain the same occurrence in the Cygnus Loop.

\citet{2013A&A...553A.104M} were the first to acquire a long-slit set of data that covered the whole SNR. Their 3.0\,\AA{} resolution spectra in the range 5950 to 6750\,\AA{} were used to describe the kinematics of the object and to map its electron density once in each $5\arcsec \times 2\arcsec$ region. This map revealed the strong density increase toward the southeast direction, which has been associated with a molecular cloud in this region \citep{1997ApJ...480..607B}.

Radiative shock models from the MAPPINGS III code were presented by \citet{2008ApJS..178...20A}. This code generates line intensity ratios for a gas in different conditions, such as the shock velocity ($100-1000$\,km\,s$^{-1}$), magnetic field strength ($10^{-4}-10.0$\,$\mu$G), and abundances (including for the LMC), and has been used to diagnose SNR properties  \citep{2010ApJ...710..964D,2012A&A...544A.140A}.

In this work, we present and analyze a set of spatially resolved data of the N\,49 SNR obtained from long-slit spectroscopy. This is the first spectroscopic mapping of a SNR in which dozens of emission lines were mapped in a wide spectral range. The data acquisition and reduction is described in Sect. \ref{obs_and_red}. Flux, flux ratio, and physical property maps are presented and discussed in Sect. \ref{results}. A general discussion and comparison with the MAPPINGS\,III model are presented in Sect. \ref{discussion}. The final conclusions are listed in Sect. \ref{conclusions}.

\section{Observations and data reduction}
\label{obs_and_red}
Spectral data of N\,49 were acquired with the 4.1\,m Southern Astrophysical Research (SOAR) telescope in Cerro Pach\'on, Chile. Observations were made on the nights of October 19, 20 and 21 and November 25, 2011. Long-slit spectroscopies were obtained by placing the slit ($\sim$$4\arcmin \times 1.03\arcsec$) over 29 different positions aligned along the east-west direction and spaced by 2$\arcsec$ from each other in declination. A 4096 $\times$ 4096 pixel Fairchild CCD and a 300\,mm$^{-1}$ grating were used, giving a spectral resolution of $\sim$\,$7$\,\AA{} at 5790\,\AA{} (resolving power $R \sim 800$) and a spectral dispersion of 1.3\,\AA{}\,pixel$^{-1}$ and a spacial scale of 0.145$\arcsec$\,pixel$^{-1}$. The spectral range was from 3550\,\AA{} to 8920\,\AA. Exposure time was 1200\,s for each slit position (one exposure each). Seeing was about $1\arcsec$ for all nights.

\begin{figure}
  \resizebox{\hsize}{!}{\includegraphics[angle=-90]{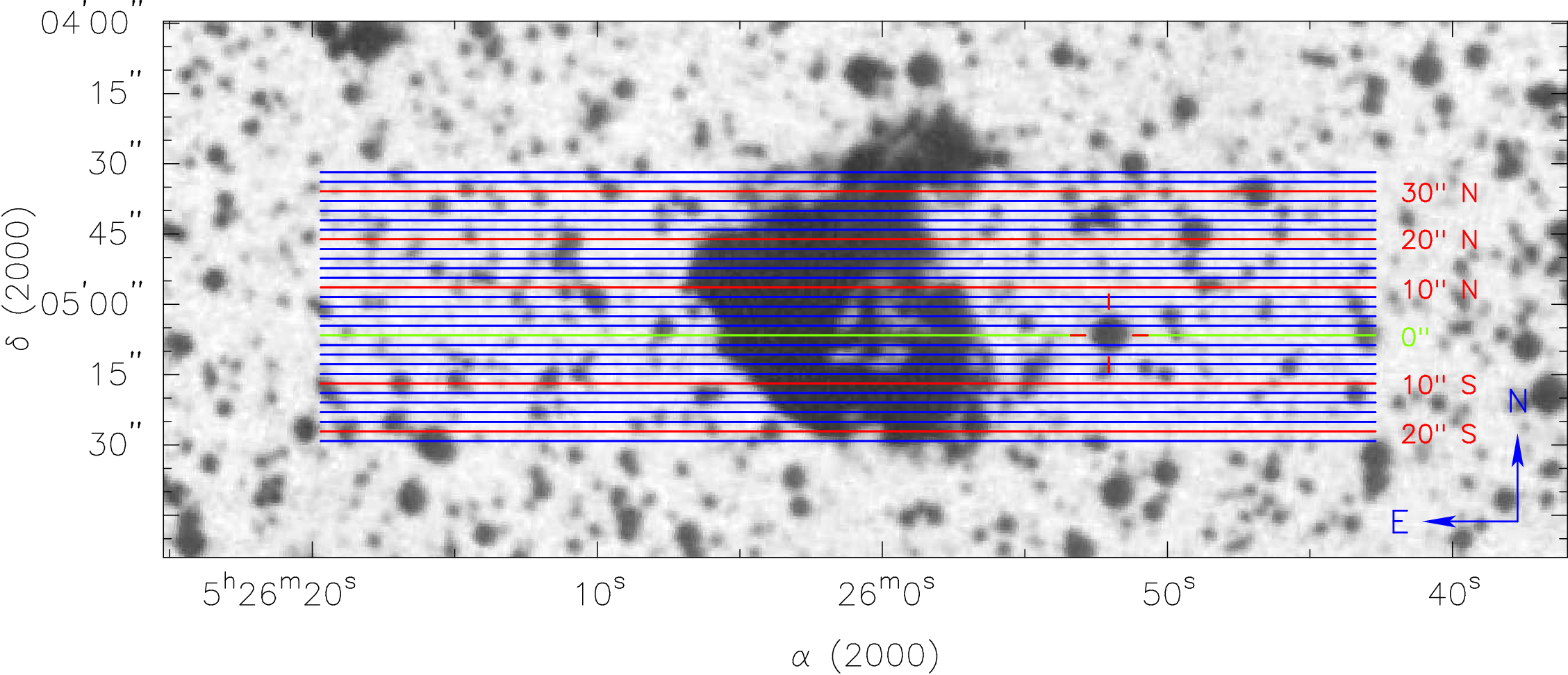}}
  \caption{Lines representing the slit positions over a J-band image of N\,49 from the SERC survey, obtained with the ALADIN software from the Centre de Donn\'ees Astronomiques de Strasbourg. The green line identifies the slit on which the reference star (rounded by red ticks) lies. Slit widths are not to scale.}
  \label{slits}
\end{figure}

The slit positions covered a field from 22$\arcsec$ south to 34$\arcsec$ north from the reference star 2MASS 0525522-6605074 ($\alpha = 05^h25^m51.6^s$, $\delta =-66\degr05\arcmin05.5\arcsec$; J2000), covering almost the entire extension of the remnant. Figure \ref{slits} shows the slit positions over a J-band image of N\,49. The reference star and the offsets for some slit positions are also indicated.

\begin{figure}
  \resizebox{\hsize}{!}{\includegraphics[angle=-90]{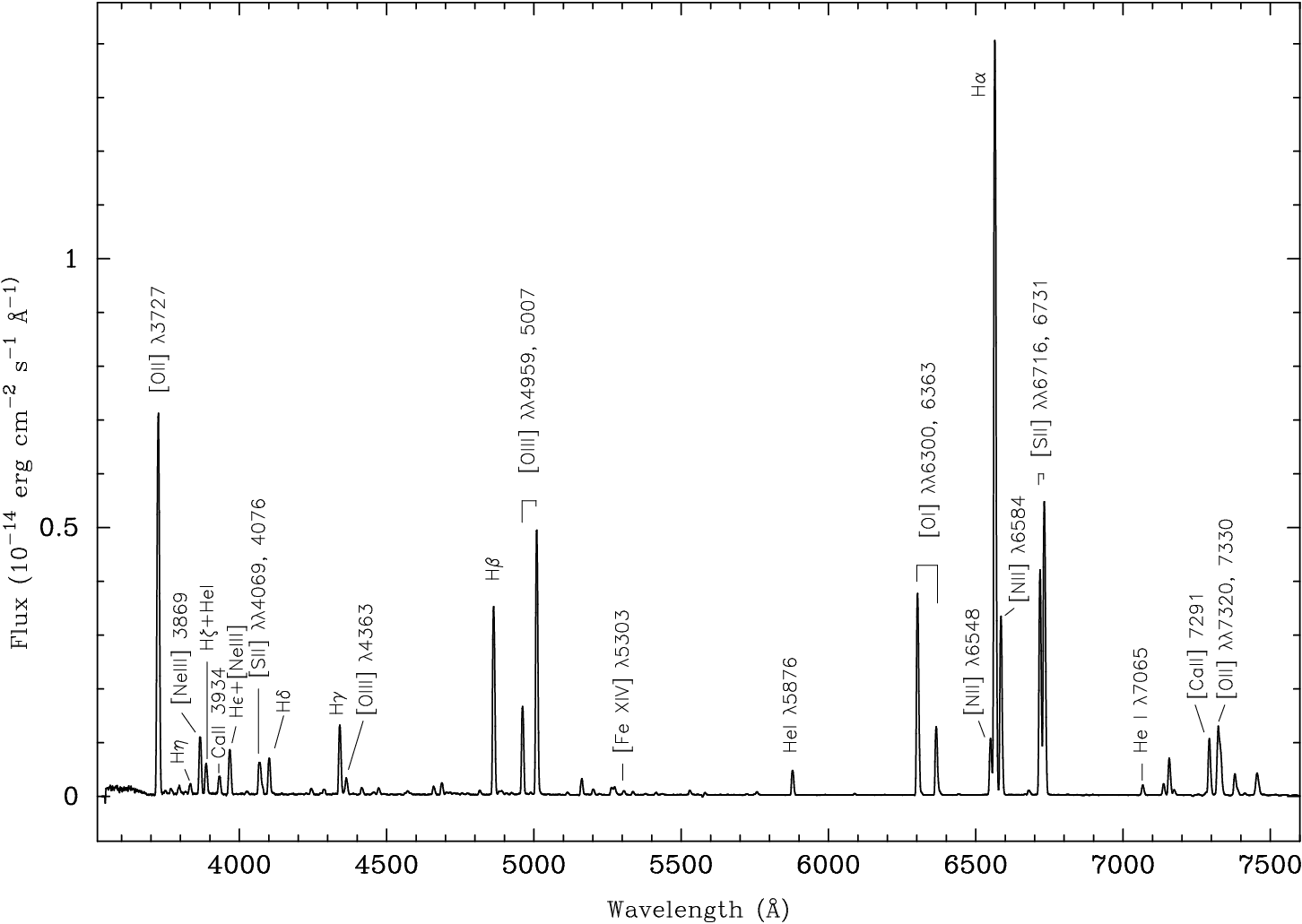}}
  \caption{Spectrum of the aperture at 74$\arcsec$ east of the reference star.}
  \label{spec}
\end{figure}

Data reduction was carried out using IRAF\footnote{IRAF is distributed by the National Optical Astronomy Observatory, which is operated by the Association of Universities for Research in Astronomy (AURA) under cooperative agreement with the National Science Foundation.}. The process primarily included overscan and bias subtraction, flat-field correction, and cleaning from cosmic-ray hits. A total of 112 apertures ($2\arcsec$ wide each) were extracted from each 29 2D spectrum. This procedure produced 3248 apertures (1D spectra). The 1D spectra were then wavelength and flux calibrated. Flux calibration was made using a sensitivity function derived from observations of the spectrophotometric standard star Feige 110 ($\alpha = 23^h19^m58.4^s$, $\delta =-05\degr09\arcmin56.2\arcsec$; J2000). These observations were taken each night with the slit aligned to the parallactic angle and widened to 3.0$\arcsec$. Sky correction was conducted by subtracting an average spectrum of about a dozen apertures that had no object emission; these were typically the easternmost apertures. Figure \ref{spec} presents the spectrum of an aperture that represents a bright portion of the remnant, 74$\arcsec$ of from the reference star. Emission line fluxes were finally measured from these spectra by Gaussian fitting of the line profile with a continuum baseline defined by eye using the \textit{splot}/IRAF task. We were unable to resolve red and blue components, therefore the measured fluxes correspond to the integrated value of both. The formal Poissonic errors in the intensities of the strongest lines are calculated to be about 1\% on the brightest areas. These estimates are clearly underestimated because errors introduced by the definition of the level of the continuum non-Gaussian form of the line and contribution of blends are neglected. Unfortunately, we were unable to estimate the errors from statistics of different measurements since there is only one exposure for each position.

Some criteria were applied to distinguish real emission lines from noise features. First, the flux must be greater than 1$\times10^{-18}$\,erg\,cm$^{-2}$\,s$^{-1}$, which is about the lower limit of the dynamic range of the measurements. Second, the feature peak value must be greater than 2.5 times the dispersion of the continuum (2.5$\sigma$) in the vicinity of the feature.

\section{Results}
\label{results}

\subsection{Flux maps}
\label{flux maps}
A total of 67 optical emission lines were measured in the final spectra. Some of the brightest emission lines (such as H$\alpha$ and [O\,{\sc iii}]\,$\lambda$5007) were present in about 2000 spectra (pixels on the maps), corresponding to positions over the remnant and beyond it. About 40 lines were mapped at least in the optical region of the remnant. 

\begin{figure}
\centering
\includegraphics[width=\linewidth]{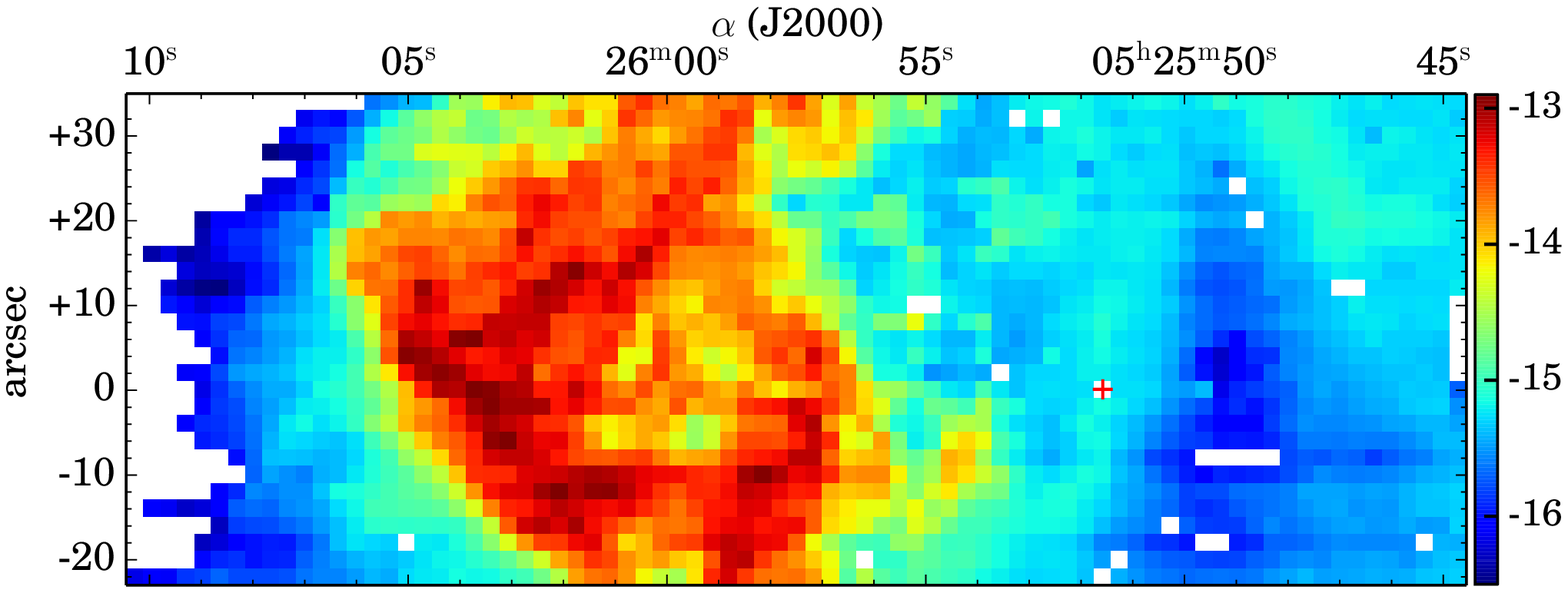}

\vspace{-1.3mm}
     
\includegraphics[width=\linewidth]{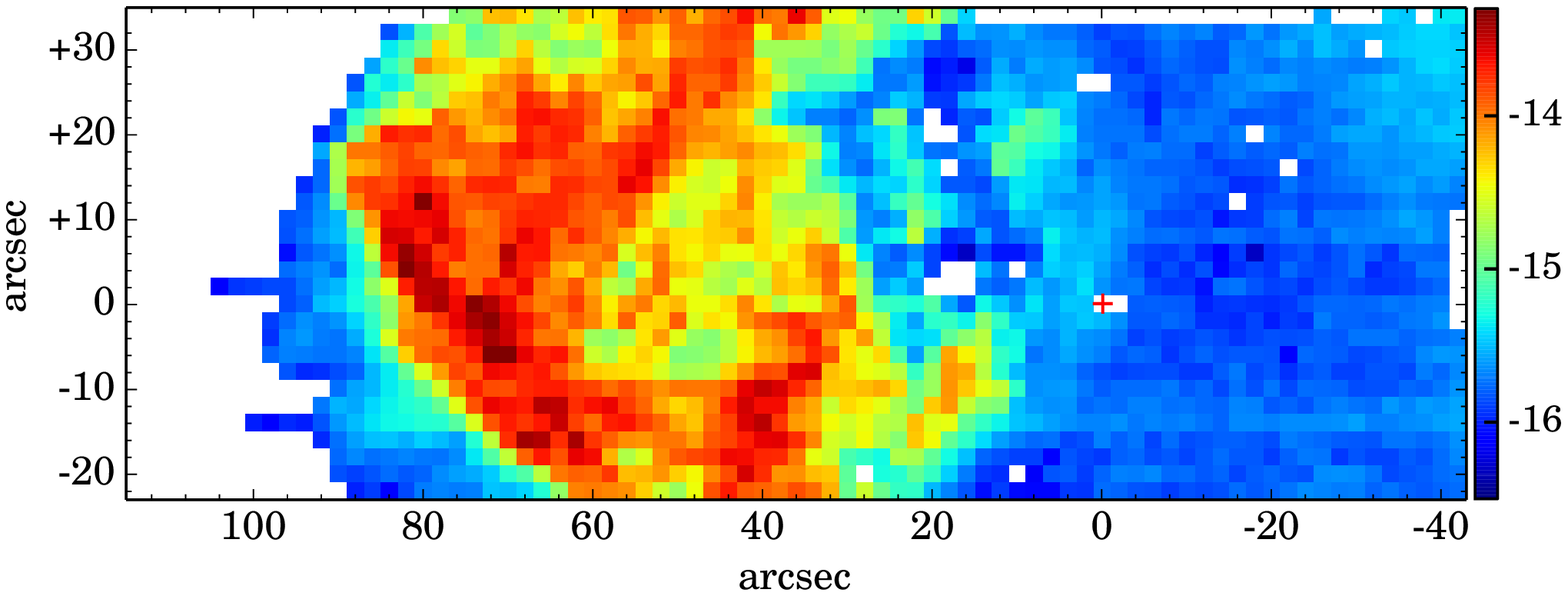}
\caption{Maps of the observed H$\alpha$ (top) and [O\,{\sc iii}]\,$\lambda$5007 (bottom) fluxes (in logarithmic scale and in units of erg\,cm$^{-2}$\,s$^{-1}$). The red cross marks the reference star position ($\alpha = 05^h25^m51.6^s$, $\delta =-66\degr05\arcmin05.5\arcsec$; J2000). Axes indicate the offsets (in arcsec) of the reference star. North is at the top and east is to the left. Each pixel represents a $2\arcsec$\,$\times$\,$2\arcsec$ region on the plane of the sky.}
\label{flux_strong}
\end{figure}

Derived H$\alpha$ and [O\,{\sc iii}]\,$\lambda$5007 observed flux maps are shown in Fig. \ref{flux_strong}. These maps show some interesting structures of the remnant, such as the central cavity (around offsets 50$\arcsec$ east and 04$\arcsec$ north) and the faint western boundaries. These boundaries were also visible in images of N\,49 presented by \citet{1992ApJ...394..158V}. The boundary in the H$\alpha$ map is farther from the remnant center than that in the [O\,{\sc iii}] map. The northwestern bright region beyond the 10$\arcsec$ western offset seen in the H$\alpha$ map is a portion of the H\,{\sc ii} region \object{DEM L 181}. White pixels represent regions without measurements, which generally means that the feature was absent or very weak, and was therefore rejected by the criteria previously explained.

Many iron lines in different ionization levels were mapped, especially [Fe\,{\sc ii}] lines. The presence of strong iron lines is a characteristic of shock-energized objects because shocks are able to destroy grains. Because of this, the iron line intensities can be up to 100 times stronger than in photoionized regions. \citet{2016arXiv160502385D} presented the results of an optical integral-field spectroscopy of N\,49. Based on emission maps of different [Fe\,{\sc ii}], [Fe\,{\sc iii}], [Fe\,{\sc v}], [Fe\,{\sc x}], and [Fe\,{\sc xiv}] lines, they have detected a segregation of the different ionic states of iron that cannot be interpreted as a mere projection effect.

\begin{figure}
\centering
\includegraphics[width=\linewidth]{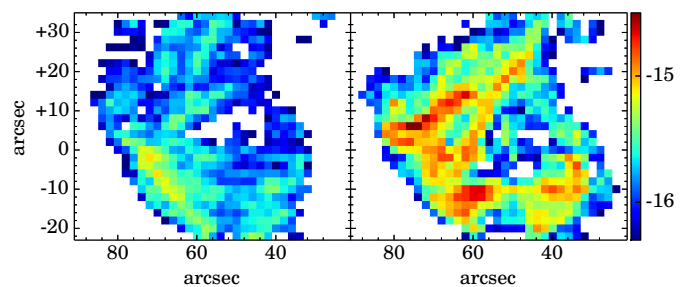}
\caption{Same as Fig. \ref{flux_strong} for [Fe\,{\sc xiv}]\,$\lambda$5303 (left) and [N\,{\sc i}]\,$\lambda$5199 (right) lines.}
\label{flux_medium}
\end{figure}

Figure \ref{flux_medium} shows [Fe\,{\sc xiv}]\,$\lambda$5303 (the highest ionization line mapped) and [N\,{\sc i}]\,$\lambda$5199 flux maps (one of the low-ionization lines measured). As expected, these maps reveals that the bright features in each map are not spatially coincident, but complementary to a certain degree. The reason is that the emission of these lines originates from very different ionization regions. Emission of [Fe\,{\sc xiv}]\,$\lambda$5303 is found throughout almost the entire extension of the remnant, indicating that fast shocks ($v_{\rm s}$\,$>$\,$360$\,km\,s$^{-1}$) are present throughout the nebula \citep{1979ApJ...231L.147D}.

\subsection{Line intensity ratio maps}
\label{line_ratios}

\begin{figure*}
\includegraphics[width=\linewidth]{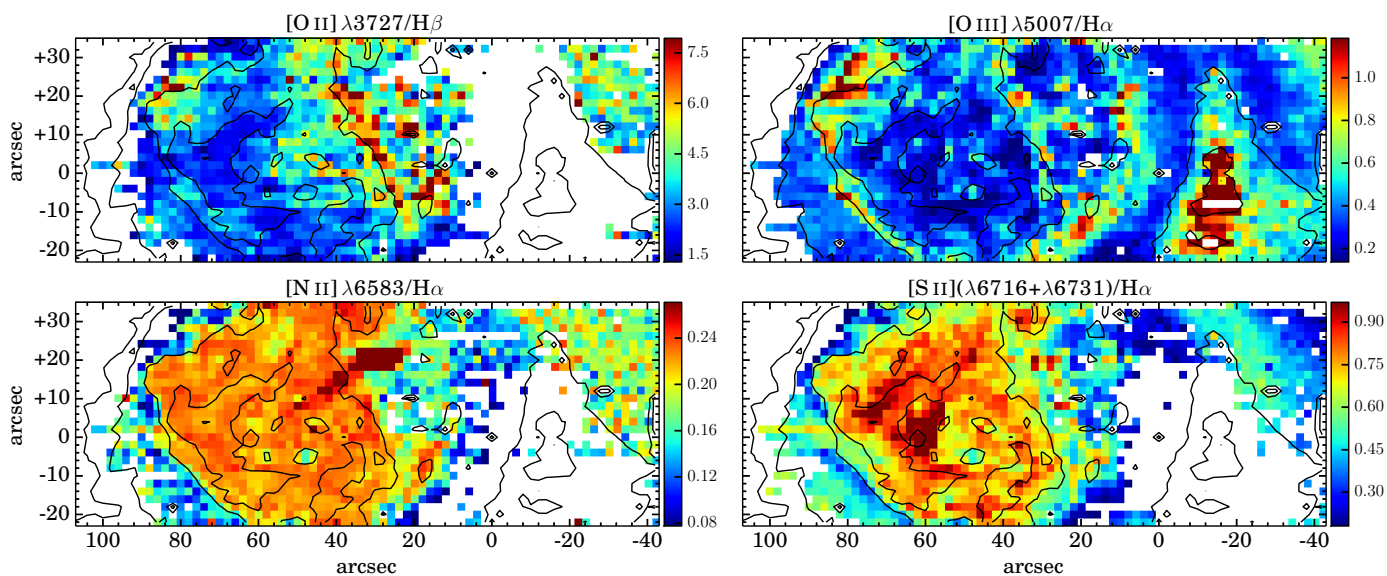}
\caption{Line ratio to H$\alpha$ or H$\beta$ maps for strong lines. Values are presented in linear scale. Contours correspond to 5.0$\times10^{-14}$, 8.0$\times10^{-15}$, 3.9$\times10^{-16}$, and 1.3$\times10^{-16}$\,erg\,cm$^{-2}$\,s$^{-1}$ from the H$\alpha$ flux map. To highlight the changes in the line ratios and mitigate outlier effects, the limits of the color scale are the 3\% and 97\% percentiles.}
\label{ratio_strong}
\end{figure*}

\begin{figure*}
\centering
\includegraphics[width=\linewidth]{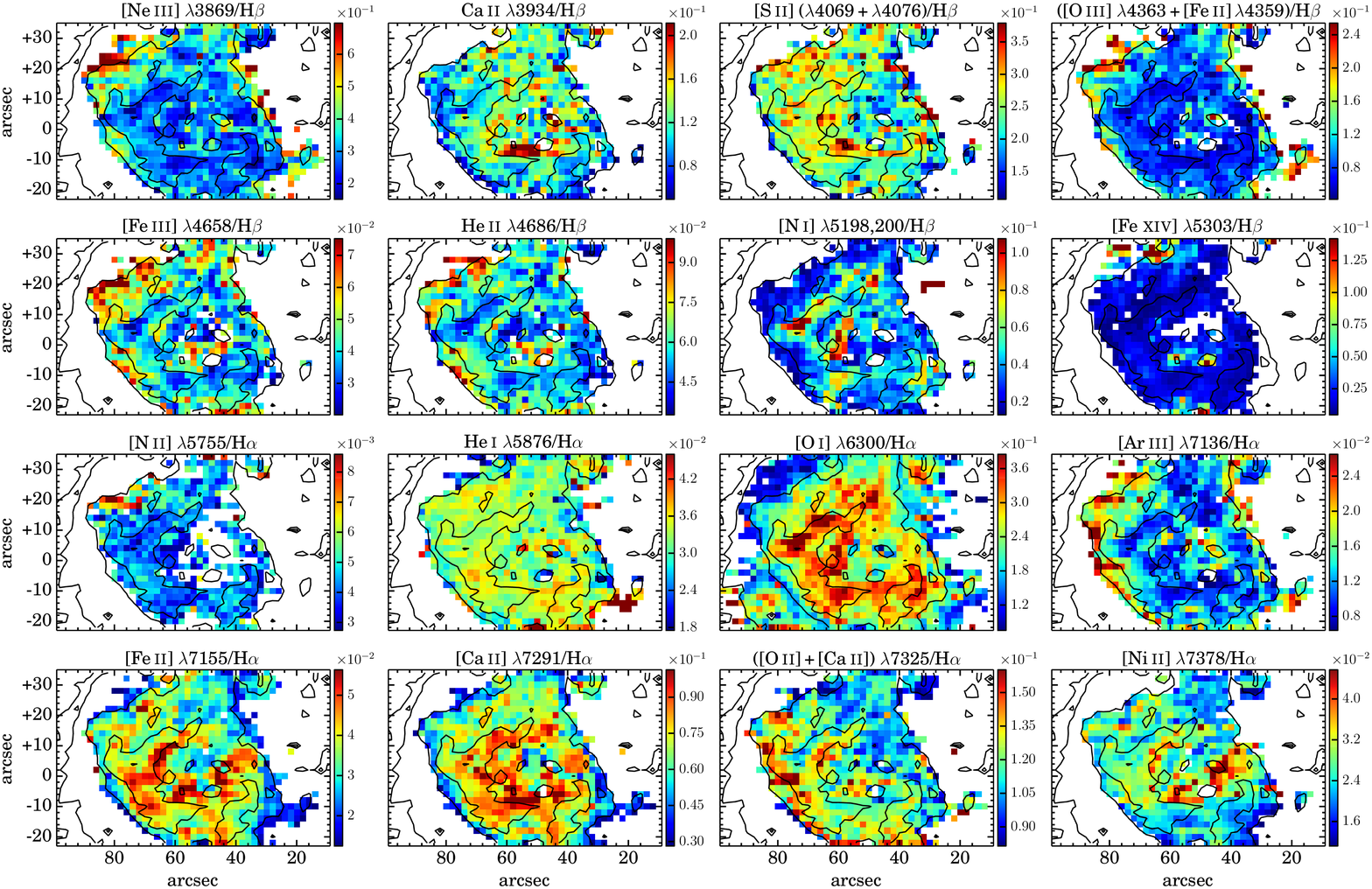}
\caption{Same as Fig. \ref{ratio_strong} for a sample of medium-intensity lines relative to H$\alpha$ or H$\beta$.}
\label{ratio_medium}
\end{figure*}

This section presents a set of line flux ratios to H$\alpha$ or H$\beta$ maps of N\,49 (Figs. \ref{ratio_strong} and \ref{ratio_medium}). Flux ratio maps are interesting because they reveal features of the observed object that are mostly unrelated with its flux maps, as was also seen by \citet{1983ApJ...273..219H} in their photometric maps of the Galactic SNR Cygnus Loop. N\,49 ratio maps were overlaid with contours of the H$\alpha$ flux to allow matching features in different maps. Many ratio maps present some organized spatial variations that certainly are related to variations of one or more physical conditions in the remnant.

\citet{1982ApJ...262..171F} analyzed ratios of flux to Balmer lines in different portions of the Cygnus Loop and found that line ratios with high-ionization lines (such as [O\,{\sc iii}], [Ne\,{\sc iii}], and [Ar\,{\sc iii}]) are correlated with each other and anticorrelated with line ratios of low-ionization lines ([O\,{\sc i}], [N\,{\sc i}], [S\,{\sc ii}], and so on). The [O\,{\sc ii}]\,$\lambda$3727/H$\beta$, [O\,{\sc iii}]\,$\lambda$5007/H$\alpha$, [N\,{\sc ii}]\,$\lambda$6583/H$\alpha$, and [S\,{\sc ii}]\,($\lambda$6716+$\lambda$6731)/H$\alpha$ ratio maps of N\,49 are presented in Fig. \ref{ratio_strong}. The first two maps show an outside ring with high values that seems to trace the limits of the remnant. In the [O\,{\sc ii}]\,$\lambda$3727/H$\beta$ ratio map, this ring is better defined at the western boundary. The ring structure is complete in the [O\,{\sc iii}]\,$\lambda$5007/H$\alpha$ map, which also shows a trail of high ratio values between the center and the northwestern region that is coincident with the contour of the remnant's brightest optical limits (between the offsets 40$\arcsec$ east and 22$\arcsec$ east). 

\begin{figure}
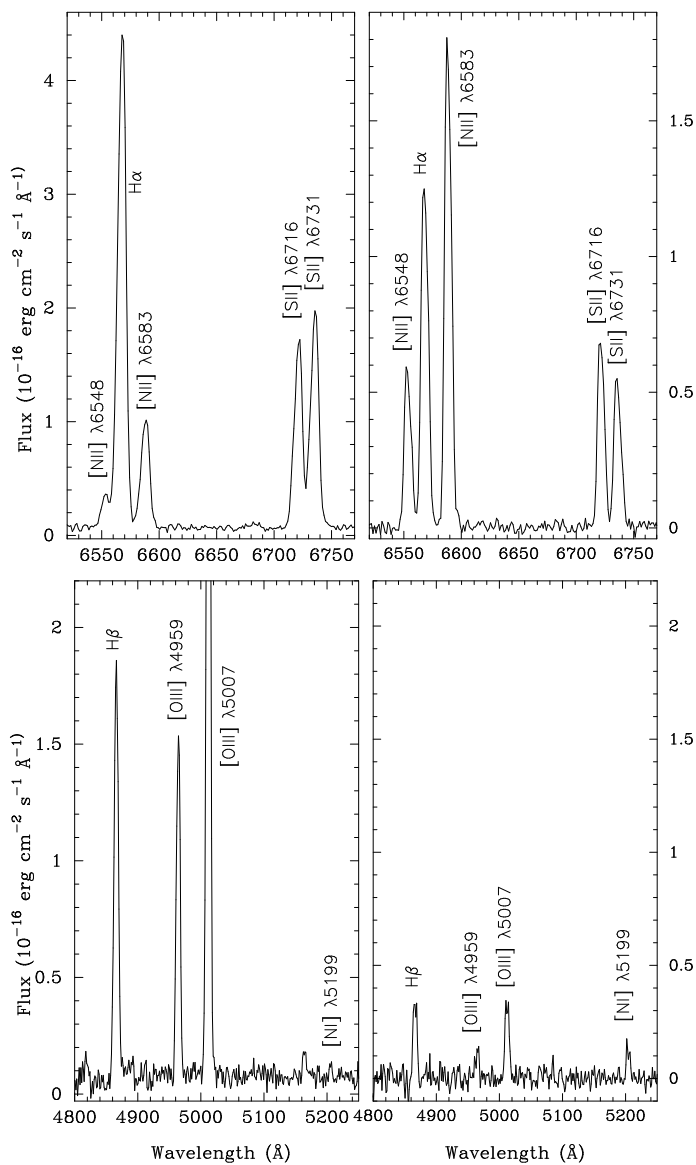

\includegraphics[height=\linewidth,angle=-90]{NII_intense.ps}
\vskip 0.2cm
\includegraphics[height=\linewidth,angle=-90]{NII_intense_5199.ps}

\caption{Sections of the spectra including the lines [N\,{\sc ii}]\,$\lambda$6583 and [N\,{\sc i}]\,$\lambda$5199 from an ordinary position in the remnant (offsets 20$\arcsec$ north and 42$\arcsec$ east, left panels) and from a position with high [N\,{\sc ii}]/H$\alpha$ (offsets 20$\arcsec$ north and 30$\arcsec$ east, right panels).}
\label{NII_intense}
\end{figure}

Another interesting region becomes prominent in the [N\,{\sc ii}]\,$\lambda$6583/H$\alpha$ map. An area with high values of this ratio starts from the SNR center and extends to its northwestern limits. The [N\,{\sc ii}]\,$\lambda$6583/H$\alpha$ values are about 0.23 at ordinary positions into the nebula, but reach values of up to 1.4 at the external edge of this area. Figure \ref{NII_intense} presents spectra from these regions. The intensities of nitrogen emission-lines are strongly correlated with the abundance of this element \citep{1995AJ....110..739L}. The [N\,{\sc i}]\,$\lambda$5199/H$\beta$ map (shown in Fig. \ref{ratio_medium} along with other selected flux ratio maps) also presents high values at exactly the same region of highest [N\,{\sc ii}]\,$\lambda$6583/H$\alpha$ values, but the inner part of the remnant in unremarkable. The lack of high [N\,{\sc i}]\,$\lambda$5199/H$\beta$ in the internal segment of the area may be caused by nitrogen being mostly ionized to N$^+$ there. However, in the outer part of this area both [N\,{\sc i}] and [N\,{\sc ii}] are intense, indicating an overabundance of nitrogen there. This nitrogen-rich area might have come from the supernova ejecta or from the wind of the pre-supernova star.

The ratio of [S\,{\sc ii}]\,($\lambda$6716+$\lambda$6731) to H$\alpha$ (Fig. \ref{ratio_strong}) presents a radial decrease from the center to the borders. The lowest values of about 0.3 are found on a ring that is coincident with the ring of upper [O\,{\sc iii}]\,$\lambda$5007/H$\alpha$ values located at the border of the SNR. Outside this ring, especially toward the southeast, the ratio [S\,{\sc ii}]/H$\alpha$ increases again. This characteristic is also observed in the [O\,{\sc i}]\,$\lambda$6300/H$\alpha$ map. The line emission from low-ionization ions such as S$^+$ and O$^0$ from this region comes from the photodissociation region associated with N\,49. Molecular emission is strong to the southeast of the SNR \citep{2012ApJ...744..160S}, indicating the presence of a molecular cloud. The area in the northwest of this ring is part of the neighbor H\,{\sc ii} region DEM L 181.

Other authors also obtained ratio maps of [O\,{\sc iii}]\,$\lambda$5007/H$\alpha$ and [S\,{\sc ii}]\,($\lambda$6716+$\lambda$6731)/H$\alpha$ for N\,49, but using imaging techniques \citep{1992ApJ...394..158V,2007AJ....134.2308B}. The main structures in each ratio map from different works, especially the radial variations, are compatible. Although the spacial resolution of our maps are lower, the spatial coverage toward the west and sensitivity are higher. The range of values obtained by \citet{2007AJ....134.2308B} for the ratio of [S\,{\sc ii}]\,($\lambda$6716+$\lambda$6731) to H$\alpha$ (0.4 to 1.4) is very similar to the obtained here (0.3 to 1.2), although our highest [O\,{\sc iii}]\,$\lambda$5007/H$\alpha$ value (1.8) is lower than that found in their higher resolution maps (up to 2.5). In this case, the lower spacial resolution of the new map may be smoothing the higher values, which are claimed by them to be related to the small offset (0.5$\arcsec$) between the emission peaks of [O\,{\sc iii}]\,$\lambda$5007 and H$\alpha$ lines compared to the pixel size (2$\arcsec$). Our new ratio maps in Fig. \ref{ratio_medium} reveal that radial variations observed on previous ratio maps are also present in many others. In general, ratio maps of low-ionization lines relative to Balmer lines show a radial gradient increasing toward the center, while high ionization line ratio maps show the opposite. A clear exception is the [Fe\,{\sc xiv}]\,$\lambda$5303/H$\beta$ map. The highest ratio values are found in regions where almost all the flux maps fade, which are some regions in the internal cavity and the northern and southern extremes of the maps, in accordance to the same ratio map obtained by \citet{1979ApJ...231L.147D}. This is expected since these lines are produced in regions that are not spatially coincident: [Fe\,{\sc xiv}]\,$\lambda$5303 is emitted from a coronal hot plasma ($T_{\rm e}\sim10^6$\,K) while the other lines come from cooler regions ($T_{\rm e}\sim10^4$\,K).

An area with high [O\,{\sc iii}]\,$\lambda$5007/H$\alpha$ is located to the southwest. The flux maps in Fig. \ref{flux_strong} show that it follows the changes of the H$\alpha$ intensities, while the [O\,{\sc iii}] map is relatively uniform there. Although this region lies outside the remnant, the western side of the remnant has very low densities. This may be allowing a radiation flux to reach this region and generating strong [O\,{\sc iii}]\,$\lambda$5007/H$\alpha$.

\subsection{Extinction}
\label{extinction}

Balmer decrements H$\alpha$/H$\beta$, H$\gamma$/H$\beta$, and H$\delta$/H$\beta$ can be used to determine the logarithmic extinction coefficient, $c$(H$\beta$), in ionized nebulae, since they are weakly influenced by other physical conditions than the extinction itself. In photoionized nebulae, the intrinsic values of these ratios are about 2.85, 0.47, and 0.26 for an electron temperature of $10^4$\,K, electron density of $\sim$\,$10^3$\,cm$^{-3}$ and a case B situation, respectively \citep{2006agna.book.....O}. When shocks take place, as in SNRs and active galactic nuclei, the collisional excitation of hydrogen Balmer lines should not be negligible and the intrinsic values of the line ratios may change. Because of its lower threshold energy, H$\alpha$ is more likely to be produced by collisions than H$\beta$ (H$\beta$ is more likely to be produced than H$\gamma$, and so on). This is why many authors used an intrinsic H$\alpha$/H$\beta$ of about 3.0 (5\% higher than the pure recombination value) when this condition is present \citep{1979ApJ...227..131S}. We used this value to obtain the $c$(H$\beta$) from the H$\alpha$/H$\beta$ ratio and the photometric intrinsic values for the other Balmer ratios, because they are less affected.

\begin{figure}
\centering
\includegraphics[width=\linewidth]{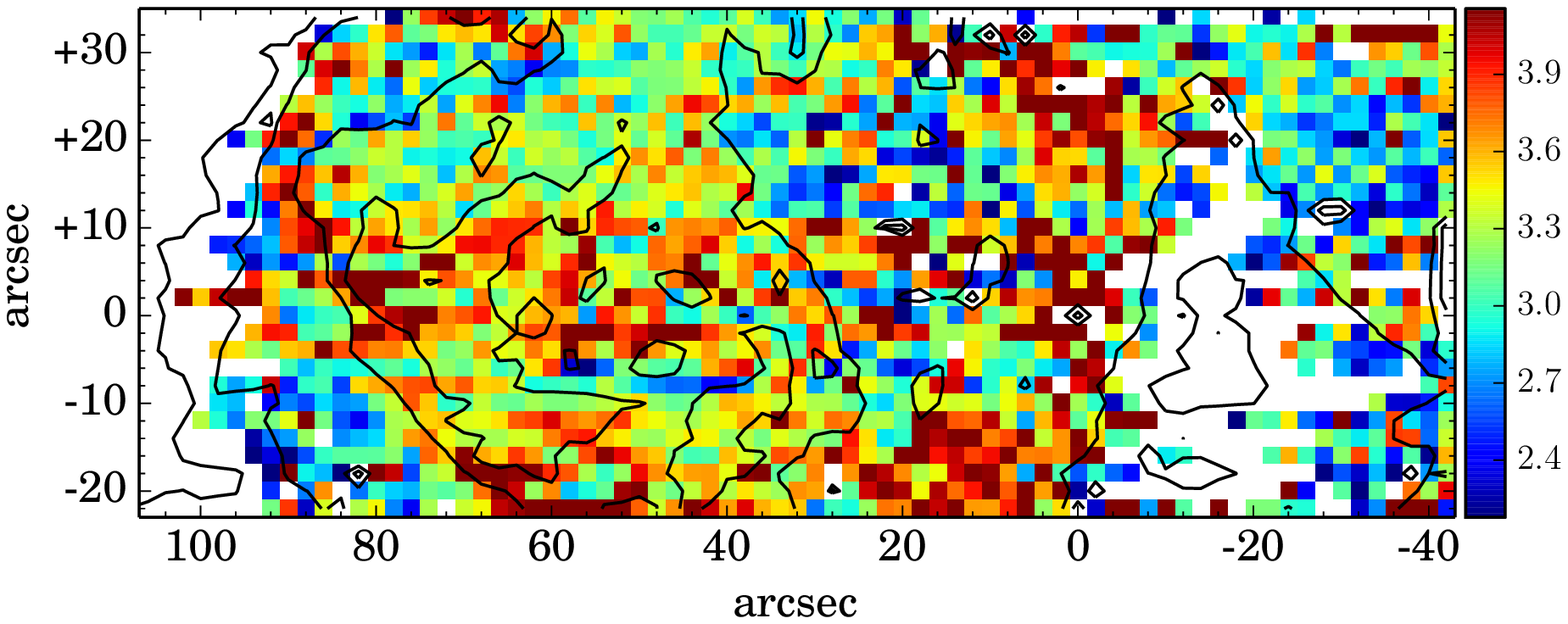}

\caption{Map of H$\alpha$ to H$\beta$.}
\label{Ha_Hb}
\end{figure}

The H$\alpha$/H$\beta$ ratio map shown in Fig. \ref{Ha_Hb} reveals an almost complete ring-like structure with higher values at the object borders (H$\alpha$/H$\beta\sim$ 4.0). A possible scenario that might generate this feature would be the existence of a high dust concentration in the remnant periphery. The dust would be pushed outward by the expanding ejected gas. This scenario is less likely because we would expect a higher extinction in the southwest region, where the molecular cloud lies. Another possibility considers additional H$\alpha$ emission that is due to collisional excitation. The high H$\alpha$/H$\beta$ ring matches the limits of the remnant reasonably well(even at the fainter western side), which may be an evidence that this ratio traces the forward shock wave front. When the H$\alpha$ line is collisionally produced, but this ignored (by assuming an intrinsic ratio H$\alpha$/H$\beta=2.85$), the result should be that any $c$(H$\beta$) estimate from the H$\alpha$/H$\beta$ ratio would be higher than the derived from the H$\gamma$/H$\beta$. Instead, the extinction values obtained from H$\alpha$/H$\beta$ (intrinsic ratio 2.85) are generally lower than those obtained from the H$\gamma$/H$\beta$ ratio (the difference is even greater if 3.0 is assumed as the intrinsic ratio). This was also observed by \citet{1986A&A...157..267D} at the same SNR, who ignored any collisional contribution to H$\alpha$ for this reason.

Models of \citet{1979ApJ...227..131S} predict a H$\alpha$/H$\beta$ ratio up to 4.2 (corresponding to a collisional excitation contribution of $\sim$30\%). Typical H$\alpha$/H$\beta$ at the borders of N\,49 are only 15\% higher than the overall mean. This shows that recombination is the chief mechanism of H$\alpha$ and H$\beta$ line emission in whole SNR, and the collisional excitation may be responsible for the increase in the H$\alpha$/H$\beta$ ratio at the borders.

\subsection{Extinction correction}
\label{deblends}
To obtain best temperature estimates, fluxes were corrected for extinction effects using results previously presented. Lines with wavelengths larger than H$\beta$ were corrected using $c$(H$\beta$) from H$\alpha$/H$\beta$. Lines with wavelengths smaller than H$\delta$ were corrected considering the $c$(H$\beta$) values obtained from H$\delta$/H$\beta$. The correction for those lines with wavelengths between H$\delta$ and H$\beta$ considered $c$(H$\beta$) values calculated from H$\gamma$/H$\beta$. When a pixel had no extinction estimate for one Balmer ratio, the $c$(H$\beta$) from the nearest Balmer line was adopted.

\subsection{Electron density and temperature}
\label{electron}

The [O\,{\sc iii}]\,$\lambda$4363 and [O\,{\sc ii}]\,$\lambda$7320,30 lines are blended with [Fe\,{\sc ii}]\,$\lambda$4359 and [Ca\,{\sc ii}]\,$\lambda$7325. We deblended them by subtracting the undesired line flux from each blend, which we calculated from other lines of the same ion for which theoretical relative intensities are known. The [Ca\,{\sc ii}]\,$\lambda$7325 line flux was obtained from the [Ca\,{\sc ii}]\,$\lambda$7291, whose relative intensities are $I(\lambda$7291)/$I(\lambda$7325)$=$3/2 \citep{1982ApJ...262..171F}. The [Fe\,{\sc ii}]\,$\lambda$4359 intensity could not be calculated directly in all the pixels where the blend [O\,{\sc iii}]+[Fe\,{\sc ii}] was measured since the [Fe\,{\sc ii}]\,$\lambda$4287  was not measured in all pixels. To solve this, we combined the spectra of these pixels (in groups of about 10) according to their ([O\,{\sc iii}]+[Fe\,{\sc ii}])/H$\beta$ value. The ([O\,{\sc iii}]+[Fe\,{\sc ii}])/[Fe\,{\sc ii}]\,$\lambda$4287 ratio of the combined spectra was used to derive the [Fe\,{\sc ii}]\,$\lambda$4287 from direct comparison with the [O\,{\sc iii}]+[Fe\,{\sc ii}] value in the pixels where [Fe\,{\sc ii}]\,$\lambda$4287 was missing. Then the [Fe\,{\sc ii}]\,$\lambda$4359 line flux map was built using the theoretical ratio of $I(\lambda$4287)/$I(\lambda$4359)$\sim$3/2 \citep{1962MNRAS.124..321G}.

Electron density and temperature maps were obtained iteratively. The density was estimated from the line ratio [S\,{\sc ii}]\,$\lambda6716/\lambda6731$  and the temperature from   [S\,{\sc ii}]\,($\lambda6716+\lambda6731$)/($\lambda4069+\lambda4076$), [O\,{\sc iii}]\,($\lambda5007$+$\lambda4959$)/$\lambda4363$, [O\,{\sc ii}]\,$\lambda3727$/$\lambda7325$, and [N\,{\sc ii}]\,($\lambda6548$+$\lambda6583$)/$\lambda5755$. The mean value over the map of one property was used to recalculate the other for pixels where the first had no value due to the lack of measured lines. Iterations were repeated until differences between pixel values in consecutive maps were smaller than 10\%.

\begin{figure}
\includegraphics[width=\linewidth]{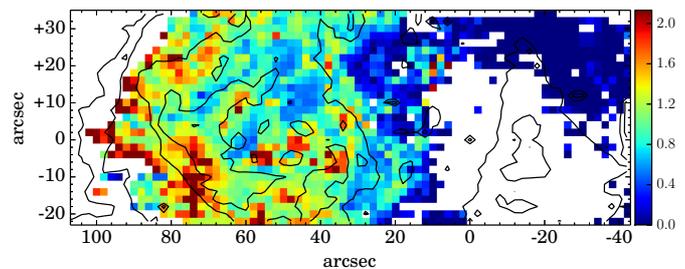}
\caption{Electron density maps in units of 10$^{3}$\,cm$^{-3}$.}
\label{dens}
\end{figure}

\begin{figure}
\includegraphics[width=\linewidth]{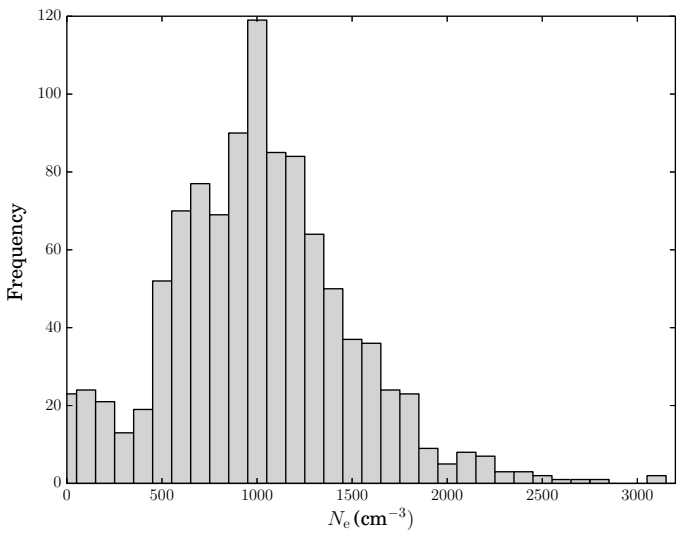}
\caption{Histogram of the electron density. The statistics used 1400 values.}
\label{histo_dens}
\end{figure}

Figure \ref{dens} shows the electron density map derived from the [S\,{\sc ii}]\,$\lambda6716/\lambda6731$ ratio. The electron density map is consistent with previous studies, especially with that from \cite{2013A&A...553A.104M}, where the increase in the density toward the southeast region was detected. The better spatial resolution and larger covered region of our data allow for a more detailed density map, however. For example, our density map shows that high-density values ($>2000$\,cm$^{-3}$) are found outside the remnant's brightest region, around offsets 92$\arcsec$ east and 4$\arcsec$ south. The overall variation in density is smooth, with a continuous increase in density from northwest to southeast, but in some positions local density peaks are found. Figure \ref{histo_dens} shows the density histogram.

\begin{figure}
\includegraphics[width=\linewidth]{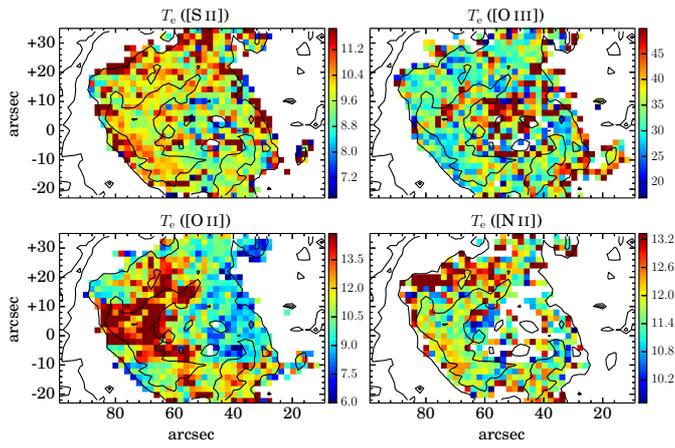}
\caption{Electron temperature maps for the [S\,{\sc ii}], [O\,{\sc iii}], [O\,{\sc ii}] and [N\,{\sc ii}] line ratios. Electron temperature unit is 10$^{3}$\,K.}
\label{temp}
\end{figure}

\begin{figure}
\includegraphics[width=\linewidth]{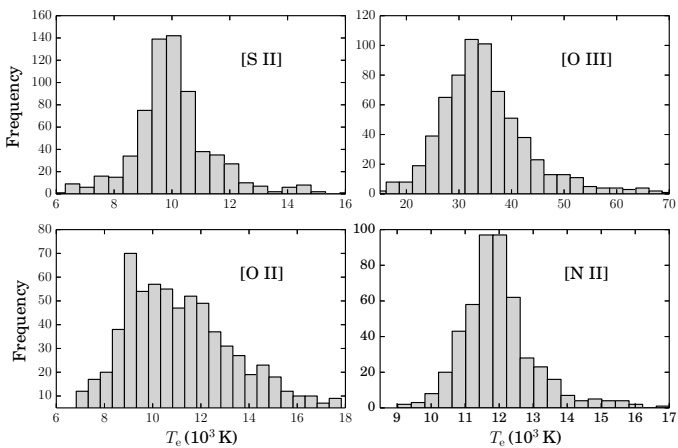}
\caption{Histograms of the electron temperature for different temperature-sensitive line ratios. The statistics for the [S\,{\sc ii}], [O\,{\sc iii}], [O\,{\sc ii}], and [N\,{\sc ii}] temperatures used 686, 710, 705, and 492 values, respectively .}
\label{histo_temp}
\end{figure}

Temperature maps from [S\,{\sc ii}]\,($\lambda6716+\lambda6731$)/($\lambda4069+\lambda4076$), [O\,{\sc iii}]\,($\lambda5007$+$\lambda4959$)/$\lambda4363$, [O\,{\sc ii}]\,$\lambda3727$/$\lambda7325$, and [N\,{\sc ii}]\,($\lambda6548$+$\lambda6583$)/$\lambda5755$ line ratios are presented in Fig. \ref{temp}. The last three temperature maps were calculated assuming the electron density values obtained in the previous process. Histograms of the electron density and temperature are presented in Fig. \ref{histo_dens} and \ref{histo_temp}. The temperature distribution throughout the remnant is different among different temperature maps. In the [S\,{\sc ii}] temperature map, values seem to increase toward the borders, while [O\,{\sc iii}] temperatures reach the highest values near the center. The [O\,{\sc ii}] electron temperature map shows higher values in the western half. Higher values of [N\,{\sc ii}] temperatures are found for some pixels at the northeastern edge of N\,49. The [N\,{\sc ii}] electron temperature map does not show values in the high [N\,{\sc ii}]\,($\lambda6548$+$\lambda6583$)/H$\alpha$ area because the [N\,{\sc ii}]\,$\lambda$5755 line is absent. The absence of [N\,{\sc ii}]\,$\lambda$5755 emission in this regions surely does not result from a low ionic abundance, since [N\,{\sc ii}]$\lambda$6583 line is intense there. Therefore, the faintness of the line indicates a low temperature in this area.

Although the temperature change in different maps is not the same, mean values are similar (about 1.1$\times10^4$\,K) for the [S\,{\sc ii}], [O\,{\sc ii}], and [N\,{\sc ii}] temperatures. Higher values are found in the [O\,{\sc iii}] temperature map, with a mean value of about 3.7$\times10^4$\,K (this value would be about 5.0$\times10^4$\,K if the [O\,{\sc iii}]\,$\lambda$4363 line would not have been deblended).

\subsection{Density and temperature estimations from the integrated spectrum compared with published data}

We combined spectra from the 800 brightest regions in the remnant to generate one spectrum with a high signal-to-noise ratio. Flux measurements from the integrated spectrum were used to obtain electron density, temperature, and extinction estimates as overall results for N\,49. Results are presented in Table \ref{integ_compar} along with dereddened flux ratio values from published data on N\,49 and properties recalculated from them. Calculation for properties from the integrated spectrum followed the same steps as those for the maps. Density values are around 10$^{3}$\,cm$^{-3}$ for most works. The higher values for \citet{1992ApJ...394..158V} and \citet{1990ApJS...74...93R} arise because the observed region is in the southwest. Our [O\,{\sc iii}] temperature value is between those determined by \citet{1992ApJ...394..158V} and \citet{1973ApJ...185..441O}. Temperatures for other line ratios are between 8.5$\times10^3$\,K to 1.3$\times10^4$\,K for all authors. Extinction results from different works do not agree perfectly. Our extinction determinations from H$\gamma$/H$\beta$ and H$\delta$/H$\beta$ are higher than those of H$\alpha$/H$\beta$. \citet{1986A&A...157..267D} data imply the same, but \citet{1990ApJS...74...93R} data show the opposite.

\begin{figure}[h!]
\resizebox{\hsize}{!}{\includegraphics{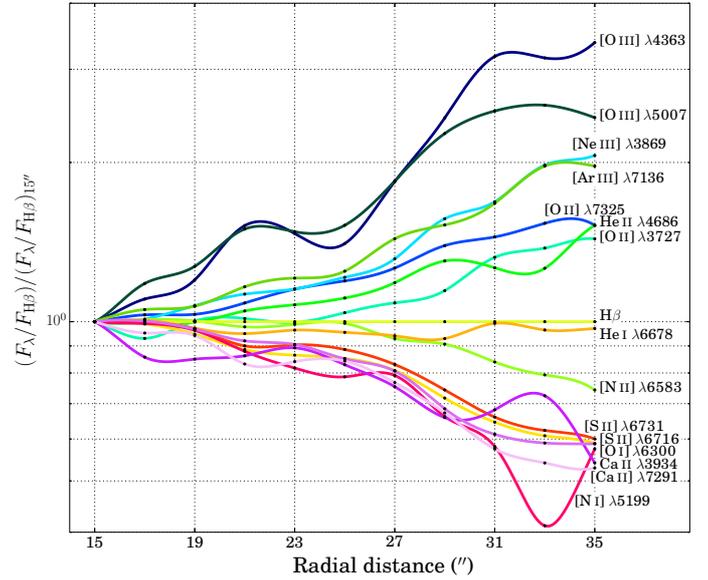}}
\caption{Profiles of line ratio maps. Points represent mean values for each line ratio (to H$\beta$) in 2$\arcsec$ concentric annuli, normalized to the value at radius 15$\arcsec$. Lines represent interpolated curves.}
\label{profiles}
\end{figure}

\begin{figure}[h!]
\resizebox{\hsize}{!}{\includegraphics{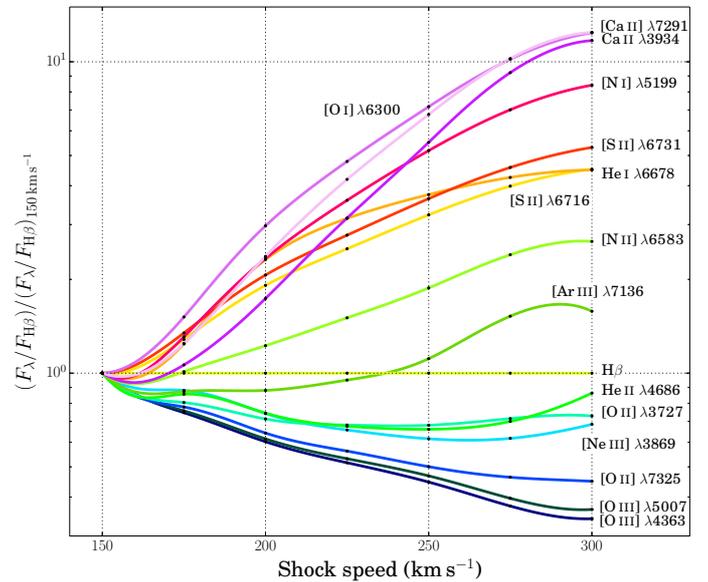}}
\caption{Line ratios to H$\beta$\,$\lambda4861$ versus shock velocity according to the MAPPINGS\,III code for LMC abundances, magnetic field strength of $10^{-4}$\,$\mu$G and an ambient density of 1\,cm$^{-3}$ for the shock component only. Points are numeric data and lines represent interpolated curves. Values are normalized to their results for $v_{\rm s}=150$\,km\,s$^{-1}$. Colors identifying each line ratio are the same as in Figure \ref{profiles}.}
\label{ratio_vel}
\end{figure}

\begin{table*}
\caption{Properties derived from the integrated spectrum and recalculated using line ratio values from published data.}
\label{integ_compar}
\centering
\begin{tabular}{r r r r r r r}
\hline
\hline
\noalign{\smallskip}
				Line ratio / property		&Integrated 	&M13 		&V92        	& R90         &D86 	  &O73$^a$ 	\\
\hline                                                                                                                                                  
\noalign{\smallskip}                                                                                                                                    
[S\,{\sc ii}]\,$\lambda6716/\lambda6731$			&0.861	        &--		&0.743       	&   0.767     &0.861	  &0.905		\\
$N_{\rm e}$([S\,{\sc ii}]) (cm$^{-3}$)				&1100	        &600-3500$^b$	&1800$^c$    	&   1600      &1000	  &800		\\
\noalign{\smallskip}                                                                                                                                    
\noalign{\smallskip}                                                                                                                                    
                                                                                                                                                        
[S\,{\sc ii}]\,$\lambda\lambda6716,31/\lambda4068,76$		&7.856	        &--		&--		&  5.876      &8.443	  &10.03		\\
$T_{\rm e}$([S\,{\sc ii}]) (K)					&9600	        &--		&--		&  10\,600    &9100 	  &8500		\\
\noalign{\smallskip}                                                                                                                			
\noalign{\smallskip}                                                                                                                                    
                                                                                                                                                        
[O\,{\sc iii}]\,$\lambda\lambda5007,4959/\lambda4363$		&19.272	        &--		&--		& 15.61       &10.49$^d$&20.80		\\
$T_{\rm e}$([O\,{\sc iii}]) (K)					&36\,700        &--		&--		& 48\,400     &>10$^5$  &33\,900	\\
\noalign{\smallskip}                                                                                                                                    
\noalign{\smallskip}                                                                                                                                    
                                                                                                                                                        
[O\,{\sc ii}]\,$\lambda3727/\lambda\lambda7320,30$		&19.646	        &--		&--		& 15.8        &18.05	  &--		\\
$T_{\rm e}$([O\,{\sc ii}]) (K)					&12\,200        &--		&--		& 12\,200     &13\,100  &--		\\
\noalign{\smallskip}                                                                                                                                    
\noalign{\smallskip}                                                                                                                                    
                                                                                                                                                        
[N\,{\sc ii}]\,$\lambda\lambda6583,84/\lambda5755$		&57.526	        &--		&55.69		&144.3        &107.0	  &--		\\
$T_{\rm e}$([N\,{\sc ii}]) (K)					&12\,600        &--		&12\,600	& 8500        &9600	&--	  	\\
\noalign{\smallskip}                                                                                                                                    
\noalign{\smallskip}                                                                                                                                    
                                                                                                                                                        
H$\alpha$/H$\beta$						&3.452 	        &--		&4.36 		&  5.97       &4.40 	  &2.96		\\
$c$(H$\alpha$/H$\beta$)$_{3.00^e}$				&0.202	        &--		&0.539		&  0.992      &0.552	  &--		\\
								                                                                                        
\noalign{\smallskip}                                                                                                                                    
\noalign{\smallskip}                                                                                                                                    
                                                                                                                                                        
H$\gamma$/H$\beta$						&0.401 	        &--		&-- 		&  0.414      &0.340 	  &0.52		\\
$c$(H$\gamma$/H$\beta$)$_{0.47^e}$				&0.502	        &--		&--		&  0.406      &0.996	  &--		\\
								                                                                                        
\noalign{\smallskip}                                                                                                                                    
\noalign{\smallskip}                                                                                                                                    
                                                                                                                                                        
H$\delta$/H$\beta$						&0.215 	        &-- 		&--		&  0.231      &0.138 	  &0.36		\\
$c$(H$\delta$/H$\beta$)$_{0.26^e}$				&0.382	        &--		&--		&  0.243      &1.237	  &--		\\  
								
\noalign{\smallskip}

\hline
\noalign{\smallskip}
\end{tabular}
\begin{minipage}[t]{0.7\textwidth}
M13: \citet{2013A&A...553A.104M}; V92: \citet{1992ApJ...394..158V}; R90: \citet{1990ApJS...74...93R};  D86: \citet{1986A&A...157..267D}, O73: \citet{1973ApJ...185..441O}.\\
$^a$Not dereddened.\\
$^b$According to authors.\\
$^c$Adopted temperature $T_{\rm e}=10\,000$\,K.\\
$^d$Not deblended from Fe$^+$ line.\\
$^e$Intrinsic values.\\

\end{minipage}
                            
\end{table*}

\section{Discussion}
\label{discussion}

\subsection{Ratio maps with radial dependence}

Many line ratios in Figs. \ref{ratio_strong} and \ref{ratio_medium} present a radial variation with values increasing toward the central region (e.g. Ca\,{\sc ii}\,$\lambda3934$/H$\beta$, [O\,{\sc i}]\,$\lambda6300$/H$\alpha$, and [S\,{\sc ii}]\,$\lambda\lambda6716,31$/H$\alpha$) or toward the borders ([Ne\,{\sc iii}]\,$\lambda3869$/H$\beta$, [O\,{\sc iii}]\,$\lambda5007$/H$\beta$, and [Ar\,{\sc iii}]\,$\lambda7136$/H$\alpha$, for example). Figure \ref{profiles} shows radial profiles of line ratio intensities obtained by averaging in 2$\arcsec$ wide concentric annuli and normalizing to the value at radius 15$\arcsec$. We investigated this behavior by ordering these lines so that the ratio of each line to the next increases toward the borders. For example, [O\,{\sc iii}]\,$\lambda5007$/H$\beta$, and [Ne\,{\sc iii}]\,$\lambda3869$/H$\beta$ maps have higher values at the borders, but the [O\,{\sc iii}]\,$\lambda5007$/[Ne\,{\sc iii}]\,$\lambda3869$ map also does. This places the [O\,{\sc iii}]\,$\lambda5007$ line before the [Ne\,{\sc iii}]\,$\lambda3869$ in this order. We concluded that the ratio of two lines L$_1/$L$_2$ is positively correlated with the distance from the center when L$_1$ comes before L$_2$ in the sequence: [O\,{\sc iii}]\,$\lambda5007$, [O\,{\sc iii}]\,$\lambda4363$, [Ar\,{\sc iii}]\,$\lambda7136$, [Ne\,{\sc iii}]\,$\lambda3869$, [O\,{\sc ii}]\,$\lambda7325$, [O\,{\sc ii}]\,$\lambda3727$, He\,{\sc ii}\,$\lambda4686$, H$\beta$\,$\lambda4861$, [N\,{\sc ii}]\,$\lambda6583$, He\,{\sc i}\,$\lambda6678$, [S\,{\sc ii}]\,$\lambda6731$, [S\,{\sc ii}]\,$\lambda6716$, [O\,{\sc i}]\,$\lambda6300$, [Ca\,{\sc ii}]\,$\lambda7291$, Ca\,{\sc ii}\,$\lambda3934$, and [N\,{\sc i}]\,$\lambda5199$. The greater the distance between two lines in this list, the stronger the variation with distance of their ratio.

A sequence of lines very similar to the one of the previous paragraph can be obtained from the MAPPINGS\,III code by analyzing how line ratios change as the shock velocity changes. Figure \ref{ratio_vel} presents interpolated curves for different line ratios (to H$\beta$) modeled by the code for a shock component. Line ratios are normalized to their values at $v_{\rm s}$=150\,km\,s$^{-1}$. This plot shows that, according to the rate of increase with the shock velocity, the considered  line ratios follow a sequence very similar to that obtained from the correlation analysis that we showed in Fig. \ref{profiles}. When the shock velocity decreases from 300\,km\,s$^{-1}$ to lower values, line ratios such as [O\,{\sc iii}]\,$\lambda5007$/H$\beta$ and [Ne\,{\sc iii}]\,$\lambda3869$/H$\beta$ increase and those of [N\,{\sc ii}]\,$\lambda5199$/H$\beta$ and Ca\,{\sc ii}\,$\lambda3934$ decrease. These changes are the same as those observed in line ratio maps with increasing distance from the remnant center, meaning that regions farther from the center would have lower velocities.

According to MAPPINGS III data, for velocities lower than $\sim$150\,km\,s$^{-1}$, some line ratios invert their variation when the shock velocity decreases. For example, [O\,{\sc iii}]\,$\lambda5007$/H$\beta$ starts to decrease when the shock velocity decreases when values are lower than $\sim$150\,km\,s$^{-1}$ and, on the other hand, [N\,{\sc ii}]\,$\lambda5199$/H$\beta$ increases in this velocity regime with decreasing shock velocity. Velocities this low may be occurring specially on the very edge of the remnant, where the sharp ring structure is seen (as those found on [O\,{\sc iii}]\,$\lambda5007$/H$\beta$ and [S\,{\sc ii}]\,($\lambda$6716+$\lambda$6731)/H$\alpha$ maps). The ring structure in each map may be indicating where the specific shock velocity on which the line ratio behavior (when the shock velocity varies) inverts is predominant.

Many line ratios have almost constant modeled values for shock velocities higher than about 300\,km\,s$^{-1}$. If shock velocity changes are indeed responsible for the radial variations in line ratio maps, then shocks slower than 300\,km\,s$^{-1}$ would be required. This is in accordance with results from \citet{1992ApJ...394..158V}, who concluded that shock velocities are mainly in the interval of 40 - 270 km\,s$^{-1}$. \citet{2007AJ....134.2308B} also reported that velocities in N\,49 are about 250\,km\,s$^{-1}$ for dense clouds and lower than 300\,km\,s$^{-1}$ for low density regions.

\citet{2013A&A...553A.104M} showed that the kinematics and morphology of N\,49 cannot be explained by a simple projection effect of a spherical shell and, surprisingly, that the gas velocity, $v$, and distance from the center, $r$, are related by $v(r)=r^{-0.9}$. However, a systematic dependence of the shock velocity on radius is not expected, even for a clumpy medium. The radial variation of line ratios is related to the time-dependency of cooling and ionization processes, and this might be the reason for the coincidence between the sequences of line ratios obtained from the radial profiles (Fig. \ref{profiles}) and from MAPPINGS curves of line strength versus shock velocity (Fig. \ref{ratio_vel}). 

\section{Conclusions}
\label{conclusions}
We presented a set of maps of line fluxes and line flux ratios for N\,49, along with derived electron density, temperature, and extinction maps. Our analysis and conclusions are summarized below.

\begin{enumerate}
  
 \item Line flux maps were built for 67 different emission lines. About 40 maps have flux measurements at least in the brightest optical region of N\,49. Many line ratio maps relative to Balmer lines presented radial variations.
  
  \item We found an area of high [N\,{\sc ii}]\,$\lambda6583$/H$\alpha$ values in this ratio map. This area extends from the center to the northwestern N\,49 border, where the high ratio values ($\sim$1.4) compared to other positions in the remnant ($\sim$0.23) are found. Since this line ratio is highly sensitive to the elemental abundance, we suggest that the intense [N\,{\sc ii}] emission probably arises from a nitrogen-rich material that is expelled from the progenitor star in the explosion or as a wind in the pre-supernova phase.
 
 \item Extinction estimates were obtained using the line ratios H$\alpha$/H$\beta$, H$\gamma$/H$\beta$, and H$\delta$/H$\beta$. A ring structure around the remnant observed in the H$\alpha$/H$\beta$ maps suggests that H$\alpha$ emission has a contribution of about 30\% from collisional excitation in this region. 
 
 \item The electron density map was built using the [S\,{\sc ii}]\,$\lambda6716/\lambda6731$ line ratio. Many high values are found beyond the brightest southwest region of N\,49. The electron temperature was obtained using [S\,{\sc ii}], [O\,{\sc ii}], [O\,{\sc iii}], and [N\,{\sc ii}]. Values from single ionized ions are about 1.1$\times10^{4}$\,K and 3.6$\times10^{4}$\,K for [O\,{\sc iii}]. Temperature variation patterns are different in maps from different temperature-sensitive line ratios.
 
 \item We investigated the variation of line ratios with distance from the remnant center. MAPPINGS III models \citep{2008ApJS..178...20A} show that different line ratios depend differently on the shock velocity. The way line ratios vary and how sensitive to velocity variations each ratio is matches our observational data if we assume that the shock speed decreases toward the remnant border. However, the required dependence of the shock velocity on radius is unexpected. The time-dependency of cooling and ionization processes may be the cause of the spatial variation in the line ratios.
 
\end{enumerate}

\begin{acknowledgements} We wish to thank the SOAR staff, in particular Tina Armond, who obtained the spectroscopic data. This work was supported by the Brazilian agencies CAPES and CNPq. \end{acknowledgements} 


\begin{thebibliography}{25}
\expandafter\ifx\csname natexlab\endcsname\relax\def\natexlab#1{#1}\fi

\bibitem[{{Alikakos} {et~al.}(2012){Alikakos}, {Boumis}, {Christopoulou}, \&
  {Goudis}}]{2012A&A...544A.140A}
{Alikakos}, J., {Boumis}, P., {Christopoulou}, P.~E., \& {Goudis}, C.~D. 2012,
  \aap, 544, A140

\bibitem[{{Allen} {et~al.}(2008){Allen}, {Groves}, {Dopita}, {Sutherland}, \&
  {Kewley}}]{2008ApJS..178...20A}
{Allen}, M.~G., {Groves}, B.~A., {Dopita}, M.~A., {Sutherland}, R.~S., \&
  {Kewley}, L.~J. 2008, \apjs, 178, 20

\bibitem[{{Banas} {et~al.}(1997){Banas}, {Hughes}, {Bronfman}, \&
  {Nyman}}]{1997ApJ...480..607B}
{Banas}, K.~R., {Hughes}, J.~P., {Bronfman}, L., \& {Nyman}, L.-A. 1997, \apj,
  480, 607

\bibitem[{{Bilikova} {et~al.}(2007){Bilikova}, {Williams}, {Chu}, {Gruendl}, \&
  {Lundgren}}]{2007AJ....134.2308B}
{Bilikova}, J., {Williams}, R.~N.~M., {Chu}, Y.-H., {Gruendl}, R.~A., \&
  {Lundgren}, B.~F. 2007, \aj, 134, 2308

\bibitem[{{Chevalier}(1977)}]{1977ARA&A..15..175C}
{Chevalier}, R.~A. 1977, \araa, 15, 175

\bibitem[{{Dennefeld}(1986)}]{1986A&A...157..267D}
{Dennefeld}, M. 1986, \aap, 157, 267

\bibitem[{{Dickel} {et~al.}(1995){Dickel}, {Chu}, {Gelino}, {Beyer}, {Burton},
  {Milne}, {Spyromilio}, {Green}, {Wilkinson}, \&
  {Junkes}}]{1995ApJ...448..623D}
{Dickel}, J.~R., {Chu}, Y.-H., {Gelino}, C., {et~al.} 1995, \apj, 448, 623

\bibitem[{{Dopita} {et~al.}(2010){Dopita}, {Blair}, {Long}, {Mutchler},
  {Whitmore}, {Kuntz}, {Balick}, {Bond}, {Calzetti}, {Carollo}, {Disney},
  {Frogel}, {O'Connell}, {Hall}, {Holtzman}, {Kimble}, {MacKenty}, {McCarthy},
  {Paresce}, {Saha}, {Silk}, {Sirianni}, {Trauger}, {Walker}, {Windhorst}, \&
  {Young}}]{2010ApJ...710..964D}
{Dopita}, M.~A., {Blair}, W.~P., {Long}, K.~S., {et~al.} 2010, \apj, 710, 964

\bibitem[{{Dopita} \& {Mathewson}(1979)}]{1979ApJ...231L.147D}
{Dopita}, M.~A. \& {Mathewson}, D.~S. 1979, \apjl, 231, L147

\bibitem[{{Dopita} {et~al.}(2016){Dopita}, {Seitenzahl}, {Sutherland}, {Vogt},
  {Winkler}, \& {Blair}}]{2016arXiv160502385D}
{Dopita}, M.~A., {Seitenzahl}, I.~R., {Sutherland}, R.~S., {et~al.} 2016, ArXiv
  e-prints

\bibitem[{{Fesen} {et~al.}(1982){Fesen}, {Blair}, \&
  {Kirshner}}]{1982ApJ...262..171F}
{Fesen}, R.~A., {Blair}, W.~P., \& {Kirshner}, R.~P. 1982, \apj, 262, 171

\bibitem[{{Garstang}(1962)}]{1962MNRAS.124..321G}
{Garstang}, R.~H. 1962, \mnras, 124, 321

\bibitem[{{Hester} {et~al.}(1983){Hester}, {Parker}, \&
  {Dufour}}]{1983ApJ...273..219H}
{Hester}, J.~J., {Parker}, R.~A.~R., \& {Dufour}, R.~J. 1983, \apj, 273, 219

\bibitem[{{Levenson} {et~al.}(1995){Levenson}, {Kirshner}, {Blair}, \&
  {Winkler}}]{1995AJ....110..739L}
{Levenson}, N.~A., {Kirshner}, R.~P., {Blair}, W.~P., \& {Winkler}, P.~F. 1995,
  \aj, 110, 739

\bibitem[{{McKee} {et~al.}(1978){McKee}, {Cowie}, \&
  {Ostriker}}]{1978ApJ...219L..23M}
{McKee}, C.~F., {Cowie}, L.~L., \& {Ostriker}, J.~P. 1978, \apjl, 219, L23

\bibitem[{{Melnik} \& {Copetti}(2013)}]{2013A&A...553A.104M}
{Melnik}, I.~A.~C. \& {Copetti}, M.~V.~F. 2013, \aap, 553, A104

\bibitem[{{Osterbrock} \& {Dufour}(1973)}]{1973ApJ...185..441O}
{Osterbrock}, D.~E. \& {Dufour}, R.~J. 1973, \apj, 185, 441

\bibitem[{{Osterbrock} \& {Ferland}(2006)}]{2006agna.book.....O}
{Osterbrock}, D.~E. \& {Ferland}, G.~J. 2006, {Astrophysics of gaseous nebulae
  and active galactic nuclei}, ed. {Osterbrock, D.~E.~\& Ferland, G.~J.}

\bibitem[{{Raymond} {et~al.}(1983){Raymond}, {Blair}, {Fesen}, \&
  {Gull}}]{1983ApJ...275..636R}
{Raymond}, J.~C., {Blair}, W.~P., {Fesen}, R.~A., \& {Gull}, T.~R. 1983, \apj,
  275, 636

\bibitem[{{Russell} \& {Dopita}(1990)}]{1990ApJS...74...93R}
{Russell}, S.~C. \& {Dopita}, M.~A. 1990, \apjs, 74, 93

\bibitem[{{Seok} {et~al.}(2012){Seok}, {Koo}, \& {Onaka}}]{2012ApJ...744..160S}
{Seok}, J.~Y., {Koo}, B.-C., \& {Onaka}, T. 2012, \apj, 744, 160

\bibitem[{{Shull} \& {McKee}(1979)}]{1979ApJ...227..131S}
{Shull}, J.~M. \& {McKee}, C.~F. 1979, \apj, 227, 131

\bibitem[{{Vancura} {et~al.}(1992){Vancura}, {Blair}, {Long}, \&
  {Raymond}}]{1992ApJ...394..158V}
{Vancura}, O., {Blair}, W.~P., {Long}, K.~S., \& {Raymond}, J.~C. 1992, \apj,
  394, 158

\bibitem[{{Williams} {et~al.}(1999){Williams}, {Chu}, {Dickel}, {Petre},
  {Smith}, \& {Tavarez}}]{1999ApJS..123..467W}
{Williams}, R.~M., {Chu}, Y.-H., {Dickel}, J.~R., {et~al.} 1999, \apjs, 123,
  467

\bibitem[{{Woltjer}(1972)}]{1972ARA&A..10..129W}
{Woltjer}, L. 1972, \araa, 10, 129

\end{thebibliography}

\end{document}